\documentclass{new_tlp}
\usepackage[utf8]{inputenc}
\usepackage[noend]{algorithmic}
\usepackage{algorithm}
\usepackage{graphicx}
\usepackage{wrapfig}
\usepackage{url}
\usepackage{subfigure}
\usepackage{multirow}

\pdfoutput=1

\title[An Or-Parallel Prolog System for Clusters of Multicores]
      {On the Implementation of an Or-Parallel Prolog System for Clusters of Multicores}

\author[J. Santos and R. Rocha]
       {JOÃO SANTOS and RICARDO ROCHA\\
       CRACS \& INESC TEC and Faculty of Sciences, University of Porto\\
       Rua do Campo Alegre, 1021, 4169-007 Porto, Portugal\\
       \email{\{jsantos,ricroc\}@dcc.fc.up.pt}}

\jdate{March 2003}
\pubyear{2003}
\pagerange{\pageref{firstpage}--\pageref{lastpage}}
\doi{S1471068401001193}

\begin{document}

\label{firstpage}

\maketitle


\begin{abstract}
  Nowadays, clusters of multicores are becoming the norm and,
  although, many or-parallel Prolog systems have been developed in the
  past, to the best of our knowledge, none of them was specially
  designed to explore the combination of shared and distributed memory
  architectures. In recent work, we have proposed a novel
  computational model specially designed for such combination which
  introduces a \emph{layered model} with two scheduling levels, one
  for workers sharing memory resources, which we named a \emph{team of
    workers}, and another for teams of workers (not sharing memory
  resources). In this work, we present a first implementation of such
  model and for that we revive and extend the YapOr system to exploit
  or-parallelism between teams of workers. We also propose a new set
  of built-in predicates that constitute the syntax to interact with
  an or-parallel engine in our platform. Experimental results show
  that our implementation is able to increase speedups as we increase
  the number of workers per team, thus taking advantage of the maximum
  number of cores in a machine, and to increase speedups as we
  increase the number of teams, thus taking advantage of adding more
  computer nodes to a cluster.
\end{abstract}

\begin{keywords}
Or-parallelism, Environment Copying, Scheduling, Implementation, Performance.
\end{keywords}


\section{Introduction}

The inherent non-determinism in the way logic programs are structured
as simple collections of alternative clauses makes Prolog very
attractive for the exploitation of \emph{implicit parallelism}. Prolog
offers two major forms of implicit parallelism: \emph{and-parallelism}
and \emph{or-parallelism}. And-Parallelism stems from the parallel
evaluation of subgoals in a clause, while or-parallelism results from
the parallel evaluation of a subgoal call against the clauses that
match that call. Arguably, or-parallel systems, such as
Aurora~\cite{Aurora-88} and Muse~\cite{Ali-90a}, have been the most
successful parallel Prolog systems so far. Intuitively, the least
complexity of or-parallelism makes it more attractive and productive
to exploit than and-parallelism, as a first step. However, practice
has shown that a main difficulty is how to efficiently represent the
\emph{multiple bindings} for the same variable produced by the
or-parallel execution of alternative matching clauses. One of the most
successful or-parallel models that solves the multiple bindings
problem is \emph{environment copying}, which has been efficiently used
in the implementation of or-parallel Prolog systems both on shared
memory~\cite{Ali-90a,Rocha-99b} and distributed
memory~\cite{Villaverde-01,Rocha-03a} architectures.

Another key problem in the implementation of a parallel system is the
design of \emph{scheduling strategies} to efficiently assign tasks to
workers. In particular, with implicit parallelism, it is expected that
the parallel system automatically identifies opportunities for
transforming parts of the computation into concurrent tasks of
parallel work, guaranteeing the necessary synchronization when
accessing shared data. For environment copying, scheduling strategies
based on \emph{dynamic scheduling of work} using \emph{or-frame data
  structures} to implement such synchronization have proved to be very
efficient for shared memory architectures~\cite{Ali-90a}. \emph{Stack
  splitting}~\cite{Gupta-99,Pontelli-06} is an alternative scheduling
strategy for environment copying that provides a simple and clean
method to accomplish work splitting among workers in which the
available work is \emph{statically divided beforehand} in
complementary sets between the sharing workers. Due to its static
nature, stack splitting was first introduced aiming at distributed
memory architectures~\cite{Villaverde-01} but, recent work, also
showed good results for shared memory architectures~\cite{Vieira-12a}.

Nowadays, the increasing availability and popularity of multicores and
clusters of multicores provides an excellent opportunity to turn
Prolog an important member of the general ecosystem of parallel
computing environments. However, although many parallel Prolog systems
have been developed in the past~\cite{Gupta-01}, most of them are no
longer available, maintained or supported. Moreover, to the best of
our knowledge, none of those systems was specially designed to explore
the combination of shared and distributed memory architectures. In
recent work, we have proposed a novel computational
model~\cite{Santos-15} which addresses the problem of efficiently
exploit the combination of shared and distributed memory
architectures. Such proposal introduces a \emph{layered model} with
two scheduling levels, one for workers sharing memory resources, which
we named a \emph{team of workers}, and another for teams of workers
(not sharing memory resources), that somehow resembles the concept of
teams used by some models combining and-parallelism with
or-parallelism, like the Andorra-I~\cite{CostaVS-91} or
ACE~\cite{Gupta-94} systems, where a layered approach also implements
different schedulers to deal with each level of parallelism.

In this work, we present a first implementation of such proposal and
for that we revive and extend the YapOr system~\cite{Rocha-99b} to
efficiently exploit parallelism between teams of workers running on
top of clusters of multicores. YapOr is an or-parallel engine based on
the environment copying model that extends the Yap Prolog
system~\cite{CostaVS-12} to exploit implicit or-parallelism in shared
memory architectures. Our platform takes full advantage of Yap's
state-of-the-art fast and optimized engine and reuses the underlying
execution environment, scheduler and part of the data structures used
to support parallelism in YapOr. In our previous
work~\cite{Santos-15}, we have described the high-level algorithms
that support the key aspects of the layered model. In this paper, we
focus the description on the operational aspects of our specific
implementation of the model, such as: how a parallel engine is
created; how a parallel goal is launched; how the team scheduler is
structured in different modules; how we deal with load balancing and
termination; how we have implemented the sharing work process using an
auxiliary sharing area; etc.

In order to take advantage of our platform, we also propose a new set
of built-in predicates that constitute the syntax to interact with an
or-parallel engine in our platform. Experimental results show that our
implementation is able to increase speedups as we increase the number
of workers per team, thus taking advantage of the maximum number of
cores in a machine, and to increase speedups as we increase the number
of teams, thus taking advantage of adding more computer nodes to a
cluster.

The remainder of the paper is organized as follows. First, we briefly
introduce the key aspects of the layered model. Next, we present the
new syntax and execution model of our platform and discuss the most
important implementation details. At the end, we present experimental
results and outline some conclusions and further work directions.


\section{Layered Model}

The goal behind the layered model is to implement the concept of
\emph{teams of workers} and specify a clean interface for
\emph{scheduling work between teams}~\cite{Santos-15}. A team is
defined as a set of workers (processes or threads) who share the same
memory address space and cooperate to solve a certain part of the main
problem. At the team level, or-parallelism can thus be explored by
using any of the available strategies tailored for
scheduling/distributing work among workers in shared memory.

\begin{wrapfigure}{r}{7.5cm}
\vspace{-\intextsep}
\centering
\includegraphics[width=7.5cm]{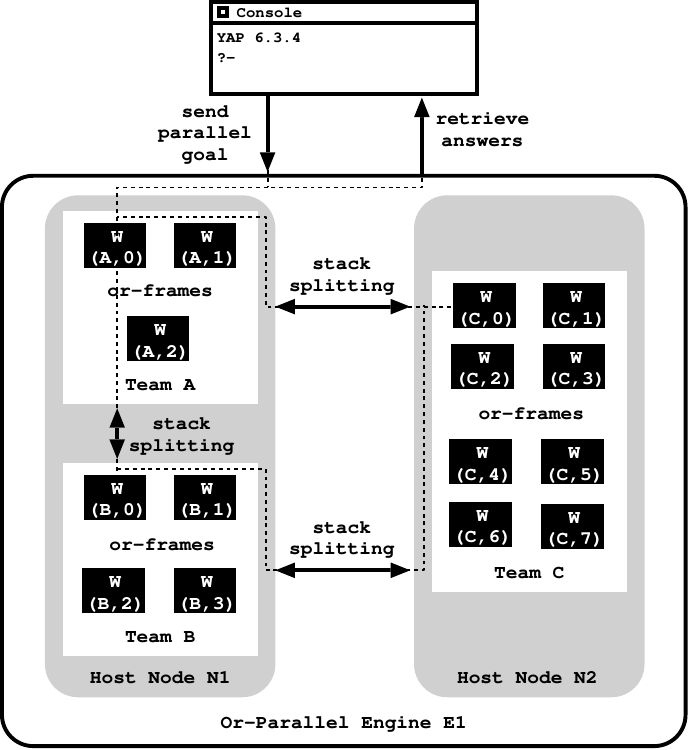}
\caption{Schematic representation of the layered model}
\label{fig_layered_model}
\vspace{-\intextsep}
\end{wrapfigure}

To schedule/distribute work between teams, the layered model
introduces a second-level team-based scheduler based on the general
ideas of the environment copying model with stack splitting. The
second-level scheduler specifies a clean and common interface for
scheduling work between teams, which fully avoids shared data
structures for synchronization by following a static work splitting
strategy among teams.

Figure~\ref{fig_layered_model} shows a schematic representation of a
possible scenario for an or-parallel engine \emph{E1} running on top
of a cluster formed by two host computer nodes. Host \emph{N1} has two
teams, team \emph{A} and team \emph{B}, with 3 and 4 workers each, and
host \emph{N2} has one team, team \emph{C}, with 8 workers. The
strategy adopted to distribute work inside each team is based on
dynamic scheduling with or-frames, while among teams is based on
static scheduling with stack splitting. This idea is similar to the
MPI/OpenMP hybrid programming pattern, where MPI is usually used to
communicate among different computer nodes and OpenMP is used to
communicate among workers in the same node. In what follows, we focus
our description on the operational aspects of our specific
implementation of the layered model.


\section{Our Platform}


\subsection{Syntax}

Our proposal for the set of built-in predicates that constitute the
syntax to interact with an or-parallel engine follows two important
design rules: (i) avoid blocking mechanisms for interacting with an
or-parallel engine, and (ii) delegate to the programmer the
responsibility of \emph{explicitly annotate} which parts of the
program should be run in an or-parallel engine in order to exploit
implicit or-parallelism. Our proposal includes five predicates:

\begin{description}
\item[par\_create\_parallel\_engine($+$EngName,~$+$ListOfTeams):]
  creates and launches the teams and workers that form a new
  or-parallel engine. The first argument defines the name to be given
  to the new or-parallel engine and the second argument is a list of
  tuples defining the teams to be created. Each tuple
  \emph{$team(h,n,p)$} includes the host computer node \emph{h} where
  the team should be launched, the number of \emph{n} workers to be
  spawned on that team and the path \emph{p} to the file program to be
  loaded by default. For example, the following call could be used to
  create the topology illustrated in Fig.~\ref{fig_layered_model}:
  \begin{center}
  \emph{par\_create\_parallel\_engine(`E1', [team(`N1',3,`prog.pl’),
      team(`N1',4,`prog.pl’), team(`N2',8,`prog.pl’)]).}
  \end{center}

\item[par\_run\_goal($+$EngName,~$+$Goal,~Template):]
  asynchronously runs a goal in an or-parallel engine. It receives as
  arguments the name of the or-parallel engine where to run the goal,
  the goal to be run and a template indicating how answers should be
  returned. Consider, for example, that we want to run the goal
  \emph{start(X,Y)} in engine \emph{E1} and we are only interested in
  the answers obtained for \emph{Y}. In such case, we should call:
  \begin{center}
  \emph{par\_run\_goal(`E1', start(X,Y), Y).}
  \end{center}

\item[par\_probe\_answers($+$EngName):] checks if the or-parallel
  engine given as argument has found new answers for the current
  parallel goal or has already finished its execution. If so, it
  succeeds. Otherwise, it fails. Using the current example would be:
  \begin{center}
  \emph{par\_probe\_answers(`E1').}
  \end{center}

\item[par\_get\_answers($+$EngName,~$+$Mode,~$-$ListOfAnswers,~$-$NumOfAnswers):]~\\
  asynchronously retrieves answers from an or-parallel engine. It
  receives as arguments the name of the or-parallel engine, how many
  answers to try to retrieve, the list where to return the answers
  found and the number of answers returned. This predicate is not
  backtrackable, however when called again it will return a new set of
  answers (if new answers exist). When all answers have been retrieved
  and the parallel goal has finished, it simply fails. The available
  options for the second argument are:
  \begin{itemize}
  \item \textbf{max(N)} -- attempts to retrieve up to \emph{N}
    answers, meaning that it will not block if the number of available
    answers is less than \emph{N}.
  \item \textbf{exact(N)} -- attempts to retrieve exactly \emph{N}
    answers, meaning that it will block while the number of answers is
    less than \emph{N} or the execution has not finished.
  \end{itemize}
  For example, if we want to retrieve exactly 10 answers from engine
  \emph{E1}, we could write:
  \begin{center}
  \emph{par\_get\_answers(`E1', exact(10), ListOfAnswers, NumOfAnswers).}
  \end{center}

\item[par\_free\_parallel\_engine($+$EngName):] terminates all the
  workers and teams for the or-parallel engine given as
  argument. Using again the current example would be:
  \begin{center}
  \emph{par\_free\_parallel\_engine(`E1').}
  \end{center}
\end{description}


\subsection{Execution Model}

At the beginning of the execution only one standard Yap engine, called
the \emph{client worker}, is running. As expected, the client worker
is responsible for interacting with the user and execute the user's
queries. In order to allow the parallel execution of goals, it is
necessary to create beforehand, at least, one or-parallel engine (our
platform allows to create several independent or-parallel engines and
run different parallel goals on each engine).

The worker 0 of each team is named the \emph{master worker} of the
team and it is responsible for launching the execution inside the team
and for the communication with the other teams. Moreover, the first
team to be launched is named the \emph{master team} and its
\emph{master worker} is also responsible for launching the execution
of the parallel goals sent by the client worker and for returning the
found answers. As described before, Fig.~\ref{fig_layered_model} shows
an example of an or-parallel engine, where we can see all these
communication links (dotted lines) between teams and with the client
worker (Yap's console on top of the figure).

If during the execution of a user's query, the client worker calls a
\emph{par\_run\_goal/3} predicate, then a message with the goal to be
run in parallel is sent from the client worker to the master worker of
the master team, worker \emph{W(A,0)} in
Fig.~\ref{fig_layered_model}. The client worker then continues its
computation, either executing other sub-computations, checking for the
existence of answers for the parallel goal, or retrieving such answers
when available. At the same time, worker \emph{W(A,0)} will notify all
the other master workers about the new received parallel goal and then
start its execution. Inside the master team, the execution behaves
like an independent YapOr engine, with the master worker sharing work
with its teammates. Outside the master team, the other teams are now
aware that a parallel computation has begun and thus they enter in
\emph{team scheduling mode}. Soon after, they will start contacting
the master team in order to get work. When a team \emph{A} receives a
sharing work request from a team \emph{B}, the \emph{team sharing work
  process} begins.

In the team sharing work process, first a worker \emph{W} from team
\emph{A} is chosen to answer the request. Then, worker \emph{W} may
reject or accept the request based on its current conditions. If the
request is accepted, \emph{W} proceeds as follows. Initially, worker
\emph{W} starts by copying its stacks to an auxiliary area assign to
itself. Then, a stack splitting strategy is applied to its stacks and
the stacks in the auxiliary area. Finally, the stacks in the auxiliary
area are sent to the master worker of the requesting team \emph{B}. In
the continuation, when the master worker of team \emph{B} receives the
stacks from team \emph{A}, they are installed on its own workspace. At
this point, the master worker of team \emph{B} informs the remaining
teammates that the team has now work. After that, the execution inside
team \emph{B} evolves as a standard YapOr execution with the master
worker performing a fail, in order to take the next open alternative,
and with its idle teammates starting to ask it for work.

A team is considered to be out of work when every worker inside the
team is idle. When that happens, the team enters again in team
scheduling mode in order to choose a busy team to request work and the
same sharing work process, as described above, is repeated. The
execution ends when all teams are idle and, in the continuation, the
client worker is notified that the parallel execution is finished.

Our platform can be seen as a layer implemented on top of Yap which
extends the existing YapOr's shared memory approach to support work
sharing between independent YapOr engines. For both layers we follow
an approach based on environment copying techniques. Our choice is
because environment copy showed to be one of the most successful
models and because YapOr already implements it as the basis for the
team concept. A team can be tough as a YapOr engine using or-frames to
synchronize the access to the open alternatives inside the team. On
top of that, a second layer responsible for distributing work among
teams was developed from scratch. This second layer supports two
different stack splitting techniques to distribute work among teams
and the communication between teams is implemented with MPI
messages.


\section{Implementation Details}


\subsection{Creating a Parallel Engine and Starting a Parallel Goal}

Predicate \emph{par\_create\_parallel\_engine/2} initiates the process
of creating an or-parallel engine. Initially, it uses the
\emph{MPI\_Comm\_spawn\_multiple()} function of the MPI API to spawn
the master workers of each team. Then, a set of MPI messages inform
each master worker about the number of workers to be present in the
corresponding team. In the continuation, each master worker allocates
the shared memory which will support the parallel execution of its
team and launches the remaining workers of the team using the
\emph{fork()} system call, in a process similar to the one in
YapOr~\cite{Rocha-99b}. Each worker will then initialize its
environment and, at last, jump to the \emph{getwork\_first\_time}
instruction. At that point, the parallel engine is ready to run.

Algorithm~\ref{alg_getwork_first_time} shows the pseudo-code for the
\emph{getwork\_first\_time} instruction. As we can see, all master
workers start by waiting for their teammates to became ready and,
after that, they guarantee that all teams are in the same condition
(i.e., all master workers wait for each other). The first wait
condition is implemented in shared memory by marking the ready workers
in a bitmap while the second wait is implemented using a MPI barrier.

\begin{algorithm}[ht]
\caption{$getwork\_first\_time(worker~W, team~T)$}
\begin{algorithmic}[1]
  \IF {$is\_master\_worker(W)$}
    \STATE $wait\_for\_teammates()$
    \STATE $wait\_for\_master\_workers()$
    \IF {$is\_master\_team(T)$}
      \STATE $msg = wait\_for\_message\_from\_client()$
      \STATE $goalTerm = get\_goal(msg)$
      \STATE $templateTerm = get\_template(msg)$
      \STATE $run\_engine(parallel(goalTerm,templateTerm))$
    \ELSE [non-master team]
      \STATE $wait\_for\_work\_in\_engine()$
      \STATE $allocate\_root\_choice\_point()$
      \STATE $team\_idle\_scheduler()$
    \ENDIF
  \ELSE [non-master worker]
    \STATE $set\_worker\_as\_ready(W)$
    \STATE $wait\_for\_work\_in\_team()$
    \STATE $allocate\_root\_choice\_point()$
    \STATE $local\_scheduler()$
  \ENDIF  
\end{algorithmic}
\label{alg_getwork_first_time}
\end{algorithm}

After this initial synchronization, the master worker of the master
team waits for a new parallel goal to be sent by the client worker
(line 5 in Alg.~\ref{alg_getwork_first_time}). Remember that predicate
\emph{par\_run\_goal/3} triggers such synchronization. When that
happens, the goal and template strings received are converted to
Prolog terms and then the parallel computation starts by running a
\emph{parallel/2} predicate with the goal and the template terms as
arguments.

The remaining master workers wait for a signal (line 10 in
Alg.~\ref{alg_getwork_first_time}) from the master worker of the
master team to be sent by the \emph{run\_engine()} procedure (line 8
in Alg.~\ref{alg_getwork_first_time}). That signal is done in a form
of a message that contains information such as the memory position of
the root choice point, that will allow all master workers to allocate
their own root choice points in the same memory position (this is
important for the environment copying technique). In the continuation,
each master worker executes the \emph{team\_idle\_scheduler()}
procedure which starts the process of looking for a busy team to
request work to (line 12 in Alg.~\ref{alg_getwork_first_time}).

On the other hand, all (non-master) workers wait for a notification
from their master worker saying that the team has work. After
receiving such notification, they also allocate their own root choice
points (line 16 in Alg.~\ref{alg_getwork_first_time}), this time such
information can be read directly, through shared memory, from the
workspace of the master worker, and then they enter in scheduling mode
in order to search for work inside the team (line 17 in
Alg.~\ref{alg_getwork_first_time}). Every time a team runs out of
work, the non-master workers return to this instruction waiting again
for its master worker to get work from another team. From their
perspective, every time the master work gets work from the outside, is
like beginning a new parallel computation.


\subsection{Team Scheduler}

The team scheduler can be divided in two different modules. The first
module, called \emph{team idle scheduler}, runs when all workers in a
team are out of work. Once it happens, the master worker of the team
enters in scheduling mode and starts running the team idle scheduler
in order to find a busy team willing to share work. The second module,
called \emph{team busy scheduler}, is run from time to time by all
workers inside a busy team and it is responsible for handling the
sharing requests sent by the idle teams. Together, both modules of the
team scheduler implement several functionalities, such as: (i) the
handling of the communications between teams; (ii) the sharing work
process between teams; and (iii) the termination process, which
determines the end of the execution of a parallel goal by ensuring
that all search space was fully explored.

\begin{figure}[ht]
\centering
\includegraphics[width=13.5cm]{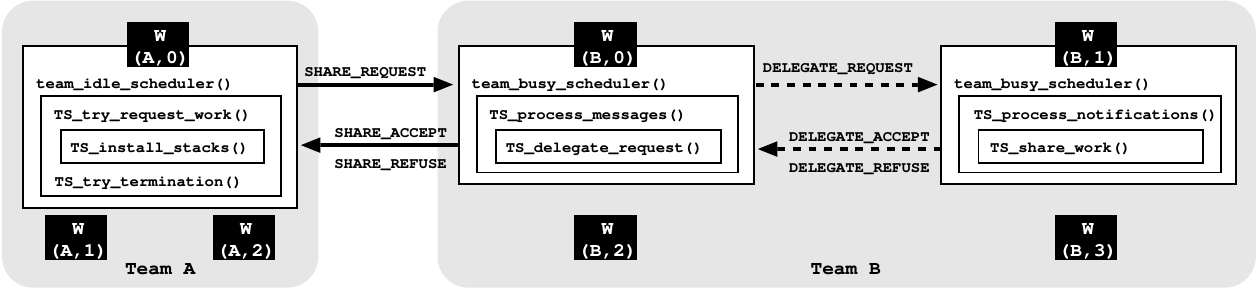}
\caption{Main components of the team scheduler}
\label{fig_team_scheduler}
\end{figure}

In order to better understand how the team scheduler works, let us
consider the example in Fig.~\ref{fig_team_scheduler}. On the left
side of the figure, we have an idle team \emph{A} formed by three
workers and, on the right side we have a busy team \emph{B} formed by
four workers. The boxes inside both teams show the main components of
the two modules of the team scheduler. The arrows represent
information exchanged during scheduler execution -- if they are
exchange between workers of the same team we call them notifications
(dotted arrows in Fig.~\ref{fig_team_scheduler}), otherwise, if they
correspond to information exchange between workers from different
teams we call them messages (solid arrows in
Fig.~\ref{fig_team_scheduler}).

Since team \emph{A} is idle, the master worker \emph{W(A,0)} starts
running the team idle scheduler and, in particular, the
\emph{TS\_try\_request\_work()} procedure responsible for selecting a
busy team to request work to. In the example, the procedure’s
algorithm decided to send a sharing request to the busy team
\emph{B}. When the master worker \emph{W(B,0)} in team \emph{B}
notices that there are pending messages coming from other teams, it
runs the \emph{TS\_process\_messages()} procedure. The request from
team A will then call the \emph{TS\_delegate\_request()} procedure in
order to select the worker inside the team with the best conditions to
successfully answer the request. In this example, the selected worker
was worker \emph{W(B,1)} and therefore a \emph{DELEGATE\_REQUEST}
notification is sent to it. Even though the sharing request involves
two teams, in practice, the sharing process is done between two
workers. The master worker of the requesting team and the chosen
sharing worker from the busy team.

Worker \emph{W(B,1)} also runs the team busy scheduler from time to
time looking for delegation requests and once it receives one, the
\emph{TS\_process\_notifications()} procedure is called. If the
request is accepted, \emph{W(B,1)} then calls the procedure
\emph{TS\_share\_work()} that is responsible for applying the stack
splitting technique, preparing the stacks to be shared and returning a
\emph{DELEGATE\_ACCEPT} notification to the master worker
\emph{W(B,0)}. Otherwise, a \emph{DELEGATE\_REFUSE} notification
informs \emph{W(B,0)} that the request was rejected. When the master
worker \emph{W(B,0)} receives the \emph{DELEGATE\_ACCEPT}
notification, it returns a \emph{SHARE\_ACCEPT} message together with
the shared stacks to the requesting team \emph{A} and, the master
worker \emph{W(A,0)} in team \emph{A}, runs the
\emph{TS\_install\_stacks()} procedure to install the received stacks
in its workspace. At this point, team \emph{A} is no longer considered
to be idle and the computation returns to Prolog execution. On the
other hand, if the sharing request is denied with a
\emph{SHARE\_REFUSE} message, team \emph{A} would try to initiate the
termination process by calling the \emph{TS\_try\_termination()}
procedure.


\subsection{Load Balancing and Termination}

An important goal of our team scheduler is to achieve an efficient
distribution of work between teams in order to optimize resource usage
and thus minimize response time. A good strategy, when searching for a
team to request work, would be to select the busy team that holds the
highest \emph{work load}, i.e., the highest amount of unexplored
alternatives. Nevertheless, selecting such a team would require having
precise information about the work load of all teams, which might not
be possible without introducing a considerable communication and
synchronization overhead. A more reasonable solution is to find a
compromise between the load balancing efficiency and the
implementation overhead.

In our implementation, each worker holds a \emph{load register}, as a
measure of the exact number of open alternatives in its private choice
points, and each team holds a \emph{load array}, as a measure of the
estimated work load of each fellow team. The load register is used to
support the process of selecting the worker with the highest load
inside the team, either when a teammate runs out of work or when
delegating a sharing request. The load array is used to support the
process of selecting a busy team to request work.

The team’s load information is stored in a bi-dimensional array which
contains, for each team, its work load and a timestamp. We define the
load of a team as the sum of the particular loads of each worker in
the team. The load array is updated whenever a team message is
received. We do not introduce specific messages to explicitly ask for
work load information and, instead, the existing messages are extended
to include that information. When a master worker sends a new message,
it includes a copy of its load array, so that the receiving master
worker can update its load array with such information. This is done
by comparing the timestamps in both arrays. When a received timestamp
is younger than the current timestamp for a given team, then the load
array is updated with the received value. Timestamps are implemented
using an integer which is incremented every time the corresponding
team sends a new message.

From a conceptual point of view, the load array can be seen as the
view that a team has about the other teams in the or-parallel
engine. Therefore, it can be useful not only for selecting teams with
work but also for detecting termination, i.e., when no busy team
exists. However, in an extreme scenario, we may have a team with load
0 but that is still busy. This may happen if, for example, all open
alternatives are in the shared region of the team. In order to
distinguish this situation, we represent the load of a team out of
work as -1, instead of 0. The termination process is thus triggered
when all teams have a load value of -1 in the load array. This is a
safe conclusion because of the way our timestamp mechanism is used to
update the load array in each team, which ensures that if a team
reaches a situation where all load values are -1 for all teams, then
the computation as ended. In such case, a termination message is
sent to all other teams signalizing that the computation has
ended. Otherwise, we restart the process of requesting work by sending
sharing requests to the workers with higher load or load 0. Meanwhile,
as other teams refuse sharing work, they will send their load array
which may contain newer information that could help to decide if the
computation has ended or not.


\subsection{Delegated Sharing Process}

When a worker receives a delegated sharing request, its stacks may not
be sent directly to the requesting team because only MPI processes can
send messages and, in our implementation, the non-master workers are
non-MPI processes since they are launched using the \emph{fork()}
system call. The \emph{TS\_delegate\_request()} procedure thus starts
by assigning an \emph{auxiliary sharing area} to each particular
delegated request. This is done by initializing a \emph{delegation
  frame data structure} associated with the request, which allows to
implement the communication between the master worker and the selected
sharing worker.

\begin{figure}[ht]
\centering
\includegraphics[width=11cm]{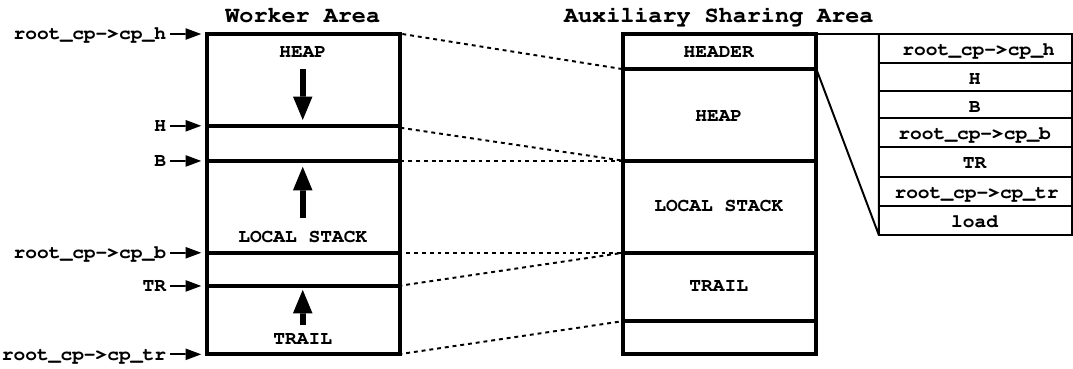}
\caption{Preparing the stacks to be sent to the requesting team}
\label{fig_preparing_stacks}
\end{figure}

The \emph{TS\_share\_work()} procedure is then responsible for
preparing the stacks to be sent to the requesting team. It starts by
determining the stacks segments to be copied. In order to better
understand how this is done, observe the left side of
Fig.~\ref{fig_preparing_stacks}, which presents the stacks in a worker
area. The area of the heap to be copied is delimited by the pointer to
the heap in the root choice point, which represents the first shared
choice point, and the register H, which points to the top of that
stack. For the local stack, the area to be copied is delimited by
register B, that points to the last choice point, and by the root
choice point. For the trail, the area to be copied is given by
register TR, that points to the top of the trail, and by the pointer
to the trail in the root choice point.

The values that delimit those areas are stored in the header region of
the auxiliary sharing area together with the load of the new team that
will be determined during the stack splitting operation.  Following
the header region are the heap, local stack and trail segments as
determined before. Since the information in the auxiliary sharing area
will be sent to the requesting team, we try to reduce the size of that
message by copying those segments in such a way that there is no gaps
between them. A schematic view of these steps is depicted on the right
side of Fig.~\ref{fig_preparing_stacks}.

Now we have two copies of the stacks, one in the workspace of the
sharing worker and another in the auxiliary sharing area, so we can
perform the stack splitting operation between both. After the stack
splitting operation completes, the load of the sharing worker and the
load in the header are both updated to reflect the changes done and
the auxiliary sharing area is ready to be sent to the requesting
team. Once the master worker of the requesting team receives it, it
just needs to install the stacks in its own workspace with the help of
the information present in the header.


\subsection{Stack Splitting}

\begin{figure}[ht]
\centering
\subfigure[Before splitting]
   {\includegraphics[height=5cm]{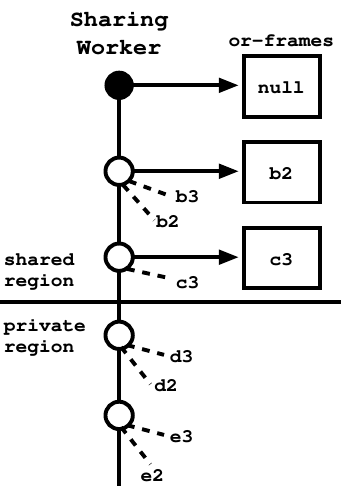}\label{fig_stack_splitting_before}}
\hspace{1.5cm}
\subfigure[After splitting]
   {\includegraphics[height=5cm]{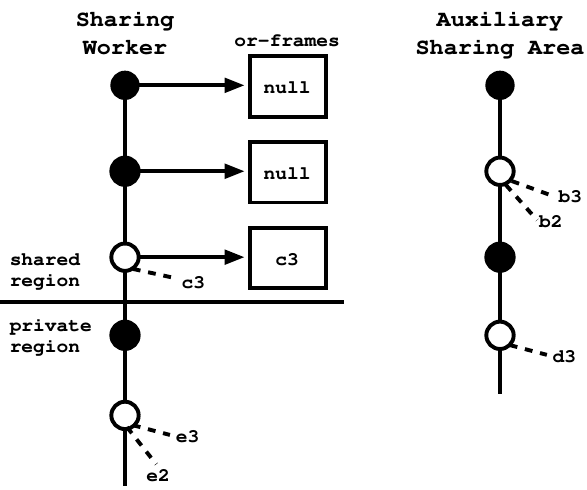}\label{fig_stack_splitting_after}}
\caption{Vertical stack splitting}
\label{fig_stack_splitting}
\end{figure}

Our platform implements two alternative stack splitting strategies for
sharing work between teams, namely \emph{vertical splitting} and
\emph{horizontal
  splitting}~\cite{Gupta-99}. Figure~\ref{fig_stack_splitting} shows a
schematic representation of the vertical stack splitting
operation.

In Fig.~\ref{fig_stack_splitting_before} we can see the execution tree
of the sharing worker which, initially, is the same as the stacks in
the auxiliary area, as explained above. The execution tree of a worker
is divided into a \emph{shared} and \emph{private} region. In the
shared region, a worker has the nodes (choice points) that are shared
with some of its teammates (nodes associated with or-frames in the
figure). On the private region, a worker has its private (non-shared)
nodes. In Fig.~\ref{fig_stack_splitting_before}, the top node is dead,
i.e., without open alternatives, but the second and third public nodes
have two (\emph{b2} and \emph{b3}) and one (\emph{c3}) open
alternatives, respectively. The two private nodes have two more open
alternatives each (\emph{d2}, \emph{d3}, \emph{e2} and \emph{e3}).

To implement vertical splitting, the sharing worker should alternately
divide its non-dead nodes between its execution stacks and the stacks
in the auxiliary sharing area. In
Fig.~\ref{fig_stack_splitting_after}, we can see the representation of
the execution tree in the sharing worker and in the auxiliary sharing
area after the vertical stack splitting operation. It is important to
note that all nodes, including the ones in the shared region, are
considered for splitting. When splitting work in a public node, first
the sharing worker needs to gain (lock) access to the corresponding
or-frame, copy the next open alternative from the or-frame to the
corresponding node in the auxiliary area, update the or-frame to
$null$ and unlock it. In the auxiliary sharing area, the or-frames
associated to the shared choice points can be simply ignored because,
after stack splitting, the work is divided in complementary sets and
such work will be installed from scratch in another team.

In the case of the horizontal splitting strategy, instead of splitting
the nodes, the sharing worker splits the open alternatives in each
node. In order to implement this strategy, a new field called
\emph{split offset} was added to the choice point and or-frame data
structures. This new field allows to calculate the alternatives
belonging to each team after a sharing process. It is initialized with
a value of one when a node is created and its value is doubled each
time the node is split with another team. When a node is turned
public, the value in the \emph{split offset} of the choice point is
simply copied to the field with the same name in the corresponding
or-frame. With the \emph{split offset}, the difference is that, when
backtracking, instead of trying the next alternative, as usual, we use
the \emph{split offset} field to calculate that alternative. For
example, if the \emph{split offset} is two, then instead of trying the
next alternative \emph{alt}, we jump \emph{alt} and we try the
alternative after \emph{alt} (offset of two).


\section{Performance Evaluation}

The environment for our experiments included two parallel machines,
each one with four AMD SixCore Opteron TM 8425 HE @ 2.1 GHz (24 cores
per machine, 48 cores in total) and 64 GBytes of main memory each,
both running Fedora 20 with the Linux kernel 3.19.8-100 64 bits. The
two machines are connected through a one Gbit router shared with other
servers. Our platform was implemented using Yap 6.3.4 and OpenMPI
1.7.3.

Since we had only available two parallel machines with the same
characteristics, we tried to emulate the existence of more computer
nodes, such that, we could create scenarios where each team of workers
always runs on a separate computer node. In order to do that, we acted
in the following way: (i) we configured OpenMPI to use the loopback
interface and the TCP protocol for communications between processes
even if they are in the same machine (by default OpenMPI uses shared
memory for this type of communications); (ii) we used the \emph{tc}
command to add more 0.06 milliseconds to the 0.02 milliseconds of
latency in the loopback interface in order to simulate the latency
that we have observed between the two physical machines, which is
about 0.08 milliseconds.

We next present a sequence of experiments using a set of 10 well-known
benchmark programs\footnote{Our platform and benchmark programs are
  available at \url{https://github.com/jpbsantos/yapor-teams}.}. We
tried to use benchmark programs widely used in the evaluation of
similar or-parallel models, but adapted to take longer when they were
too small. All together, the 10 benchmarks take around 1800 seconds
(30 minutes) to run with YapOr with a single worker.

Table~\ref{table_experiments} shows the speedup results achieved for
16, 24 and 32 workers with different configurations of teams when
compared with the execution of YapOr with a single worker. We have not
collected results for more than 16 workers per machine (32 workers in
total) because we do not had full access to the machines and since
other users could be using the machines simultaneously, thus
interfering with our results, we decided to be safer to go only until
32 workers. All teams are created with the same number of workers and
divided equitably between the two physical machines. For example,
consider the case of 24 workers and 4 teams, then we launch two teams
in each machine and each team is created with six workers each.

The results presented in Table~\ref{table_experiments} are the average
of 10 runs. The table is divided in three parts: execution with 16, 24
and 32 workers. The columns represent the configurations of teams
tested where the number of teams in which workers are divided
increases from left to right. For example, configuration
\textbf{[8,8]} means 2 teams with 8 workers each and configuration
\textbf{[4,4,4,4]} means 4 teams with 4 workers each. For each
configuration, we show the results for both vertical and horizontal
splitting, respectively, columns \textbf{VS} and \textbf{HS}. The only
exception is the case of 16 workers where we show the results for a
single YapOr engine (configuration \textbf{[16]}). For this
configuration, in parentheses, we also show the execution times in
milliseconds for YapOr with a single worker, which were used as the
base times to compute the speedups for all configurations in
Table~\ref{table_experiments}.

\begin{table}[ht]
\centering
\caption{Speedup results against YapOr execution with a single worker
  for different configurations of teams with 16, 24 and 32 workers
  using vertical (VS) and horizontal (HS) splitting on a set of 10
  well-known benchmark programs}
\label{table_experiments}
\scriptsize
\begin{tabular}{l|rr|rr|rr|rr|rr}       
\hline
\multicolumn{1}{l}{\textbf{16 Workers}} &
   \multicolumn{2}{c}{\textbf{[16]}} & 
   \multicolumn{2}{c}{\textbf{[8,8]}} &
   \multicolumn{2}{c}{\textbf{[4,4,4,4]}} &
   \multicolumn{2}{c}{\textbf{[2,2,...,2]}} &
   \multicolumn{2}{c}{\textbf{[1,1,...,1]}} \\
\hline
\textbf{Program} &
   \multicolumn{2}{c|}{\textbf{YapOr}} &
   \multicolumn{1}{c}{\textbf{VS}} & \multicolumn{1}{c|}{\textbf{HS}} &
   \multicolumn{1}{c}{\textbf{VS}} & \multicolumn{1}{c|}{\textbf{HS}} &
   \multicolumn{1}{c}{\textbf{VS}} & \multicolumn{1}{c|}{\textbf{HS}} &
   \multicolumn{1}{c}{\textbf{VS}} & \multicolumn{1}{c}{\textbf{HS}} \\
\emph{arithmetic\_puzzle} & \multicolumn{2}{r|}{ 3.39           (361.439)} &  6.10 &  6.08 &  8.07 &  8.04 &  9.91 & 10.20 &  8.31 &  7.79 \\
\emph{cubes}              & \multicolumn{2}{r|}{15.87 (\phantom{0}67.503)} & 14.29 & 13.31 & 12.94 & 12.65 &  9.82 & 10.43 &  5.31 &  6.45 \\
\emph{ham}                & \multicolumn{2}{r|}{16.01           (121.444)} & 14.17 & 14.15 & 13.49 & 13.74 &  6.50 & 11.60 &  6.92 &  6.39 \\
\emph{knight\_move(13)}   & \multicolumn{2}{r|}{15.70           (395.602)} & 15.36 & 14.94 & 14.86 & 14.73 & 13.33 & 13.30 & 10.51 & 10.24 \\
\emph{magic\_cube}        & \multicolumn{2}{r|}{15.78 (\phantom{0}46.872)} & 12.03 & 13.95 & 11.26 & 12.73 &  9.00 & 11.77 &  4.76 &  7.32 \\
\emph{map\_colouring}     & \multicolumn{2}{r|}{15.81           (178.729)} & 15.17 & 14.72 & 14.54 & 14.06 & 12.23 & 12.15 &  7.11 &  7.17 \\
\emph{nsort(12)}          & \multicolumn{2}{r|}{15.89           (406.292)} & 14.24 & 15.48 & 13.91 & 14.76 & 12.98 & 13.75 & 10.23 &  9.91 \\
\emph{puzzle4x4}          & \multicolumn{2}{r|}{15.76 (\phantom{0}17.538)} & 11.61 & 12.56 & 10.02 &  9.94 &  7.36 &  7.62 &  3.36 &  4.56 \\
\emph{queens(14)}         & \multicolumn{2}{r|}{16.09           (552.275)} & 15.00 & 15.46 & 14.33 & 15.03 & 11.88 & 14.10 &  8.09 & 11.23 \\
\emph{send\_more}         & \multicolumn{2}{r|}{15.75 (\phantom{0}69.684)} & 14.02 & 13.97 & 13.02 & 13.59 &  9.87 & 12.60 &  5.82 &  8.36 \\
\hline
\textbf{Average}          & \multicolumn{2}{c|}{14.60} & 13.20 & 13.46 & 12.64 & 12.93 & 10.29 & 11.75 &  7.04 &  7.94 \\
\hline\hline
\multicolumn{1}{l}{\textbf{24 Workers}} &
   \multicolumn{2}{c}{\textbf{[12,12]}} &
   \multicolumn{2}{c}{\textbf{[6,6,6,6]}} &
   \multicolumn{2}{c}{\textbf{[4,4,...,4]}} &
   \multicolumn{2}{c}{\textbf{[2,2,...,2]}} &
   \multicolumn{2}{c}{\textbf{[1,1,...,1]}} \\
\hline
\textbf{Program} &
   \multicolumn{1}{c}{\textbf{VS}} & \multicolumn{1}{c|}{\textbf{HS}} &
   \multicolumn{1}{c}{\textbf{VS}} & \multicolumn{1}{c|}{\textbf{HS}} &
   \multicolumn{1}{c}{\textbf{VS}} & \multicolumn{1}{c|}{\textbf{HS}} &
   \multicolumn{1}{c}{\textbf{VS}} & \multicolumn{1}{c|}{\textbf{HS}} &
   \multicolumn{1}{c}{\textbf{VS}} & \multicolumn{1}{c}{\textbf{HS}} \\
\emph{arithmetic\_puzzle} &  6.65 &  6.56 & 10.52 & 10.41 & 12.02 & 12.10 & 11.69 & 11.72 &  8.33 &  8.36 \\
\emph{cubes}              & 20.36 & 19.09 & 18.61 & 17.61 & 16.66 & 16.59 & 11.35 & 12.00 &  5.54 &  6.83 \\
\emph{ham}                & 20.83 & 20.49 & 19.81 & 20.06 & 16.86 & 18.25 &  6.66 & 14.33 &  7.01 &  6.47 \\
\emph{knight\_move(13)}   & 22.92 & 22.06 & 22.07 & 21.54 & 21.12 & 20.49 & 17.59 & 17.02 & 12.68 & 11.50 \\
\emph{magic\_cube}        & 17.26 & 20.51 & 16.20 & 18.09 & 14.52 & 17.04 &  9.43 & 13.54 &  4.11 &  7.34 \\
\emph{map\_colouring}     & 22.39 & 20.91 & 20.79 & 20.29 & 20.11 & 18.58 & 14.07 & 13.88 &  7.44 &  7.29 \\
\emph{nsort(12)}          & 21.02 & 22.42 & 20.28 & 21.58 & 19.42 & 20.82 & 16.78 & 17.92 & 11.52 &  9.88 \\
\emph{puzzle4x4}          & 15.34 & 16.85 & 13.00 & 13.78 & 11.34 & 11.83 &  7.71 &  8.18 &  3.49 &  4.76 \\
\emph{queens(14)}         & 21.97 & 23.09 & 21.30 & 22.48 & 20.07 & 21.83 & 13.71 & 20.06 &  9.95 & 12.74 \\
\emph{send\_more}         & 19.67 & 18.84 & 19.04 & 18.97 & 16.91 & 18.05 & 12.51 & 14.84 &  5.79 &  8.52 \\
\hline
\textbf{Average}          & 18.84 & 19.08 & 18.16 & 18.48 & 16.90 & 17.56 & 12.15 & 14.35 &  7.59 &  8.37 \\
\hline\hline                                                                          
\multicolumn{1}{l}{\textbf{32 Workers}} &
   \multicolumn{2}{c}{\textbf{[16,16]}} &
   \multicolumn{2}{c}{\textbf{[8,8,8,8]}} &
   \multicolumn{2}{c}{\textbf{[4,4,...,4]}} &
   \multicolumn{2}{c}{\textbf{[2,2,...,2]}} &
   \multicolumn{2}{c}{\textbf{[1,1,...,1]}} \\
\hline
\textbf{Program} &
   \multicolumn{1}{c}{\textbf{VS}} & \multicolumn{1}{c|}{\textbf{HS}} &
   \multicolumn{1}{c}{\textbf{VS}} & \multicolumn{1}{c|}{\textbf{HS}} &
   \multicolumn{1}{c}{\textbf{VS}} & \multicolumn{1}{c|}{\textbf{HS}} &
   \multicolumn{1}{c}{\textbf{VS}} & \multicolumn{1}{c|}{\textbf{HS}} &
   \multicolumn{1}{c}{\textbf{VS}} & \multicolumn{1}{c}{\textbf{HS}} \\
\emph{arithmetic\_puzzle} & 6.70  &  6.63 & 11.07 & 10.61 & 15.47 & 15.14 & 13.45 & 12.69 &  8.94 &  8.79 \\
\emph{cubes}              & 26.04 & 22.99 & 23.59 & 21.27 & 19.27 & 19.65 & 11.72 & 12.63 &  5.61 &  7.27 \\
\emph{ham}                & 26.18 & 26.65 & 25.31 & 26.28 & 20.13 & 21.52 &  6.91 & 15.36 &  7.52 &  7.42 \\
\emph{magic\_cube}        & 21.65 & 26.03 & 20.28 & 22.92 & 15.94 & 20.81 & 10.02 & 14.20 &  4.57 &  7.85 \\
\emph{knight\_move(13)}   & 30.08 & 28.10 & 28.97 & 28.16 & 26.81 & 26.16 & 20.30 & 18.09 & 13.39 & 12.27 \\
\emph{map\_colouring}     & 29.00 & 25.75 & 27.46 & 26.14 & 23.26 & 21.96 & 14.72 & 14.29 &  7.88 &  7.16 \\
\emph{nsort(12)}          & 27.80 & 28.38 & 26.50 & 27.23 & 24.07 & 25.05 & 17.56 & 18.03 & 11.55 & 11.24 \\
\emph{puzzle4x4}          & 19.95 & 21.21 & 15.25 & 17.73 & 13.77 & 13.21 &  7.77 &  8.54 &  3.40 &  5.05 \\
\emph{queens(14)}         & 28.67 & 30.62 & 27.55 & 29.42 & 23.99 & 27.99 & 15.59 & 23.64 &  8.74 & 13.29 \\
\emph{send\_more}         & 24.91 & 22.99 & 23.86 & 23.74 & 19.93 & 22.55 & 12.80 & 15.98 &  5.45 &  8.23 \\
\hline
\textbf{Average}          & 24.10 & 23.94 & 22.98 & 23.35 & 20.26 & 21.40 & 13.08 & 15.35 &  7.70 &  8.86 \\
\hline
\end{tabular}
\end{table}

Please note that most of the configurations in
Table~\ref{table_experiments} were not possible without our
platform. Before our work, parallelism could only be explored using
either shared memory based models or distributed memory based
models. With shared memory based models, we are limited by the number
of cores in the shared memory architecture. With distributed memory
based models, we can easily increase the number of available cores,
but we are limited by the potential communication and synchronization
costs as we increase the number of workers. For example, consider the
scenario where we have two multicore machines with 16 cores each (32
cores in total). With shared memory based models, we can only take
advantage of the maximum number of cores in one machine (i.e.,
configuration \textbf{[16]} in Table~\ref{table_experiments}). With
distributed memory based models, we can exploit the 32 available cores
but the potential benefits of having shared memory resources in the
multicore machines are mostly wasted. In
Table~\ref{table_experiments}, the configurations \textbf{[1,1,...,1]}
with a single worker per team match the distributed memory based model
implementing the original stack splitting approach~\cite{Gupta-99}. We
can thus see the results obtained for configurations
\textbf{[1,1,...,1]} as a way to measure and compare the former
distributed memory based models with the newer teams' configurations.

Analyzing the general picture of Table~\ref{table_experiments}, one
can observe that regarding the strategy used to distribute work among
teams, our experiments seem to suggest that, on average, horizontal
splitting achieves slightly better speedup results than vertical
splitting and that the difference between both seems to increase as we
increase the number of teams (thus also increasing the number of
potential stack splitting operations).

Reading the results in Table~\ref{table_experiments} horizontally
(i.e., considering a fixed total number of workers), the speedup
results clearly increase as we increase the number of workers per team
(thus reducing the number of teams). The exception is the
\emph{arithmetic\_puzzle} program which seems to benefit from the
higher number of stack splitting operations when we have more
teams. In this regard, the worst results are for configurations
\textbf{[1,1,...,1]}, which clearly shows the potential of our
approach of grouping workers in teams.

Reading the results in Table~\ref{table_experiments} vertically (i.e.,
considering a fixed number of workers per team), the speedup results
clearly increase as we increase the number of teams. This mimics the
scenario where we add more computer nodes with the same
characteristics to our cluster. It is interesting to note that the
speedup results obtained for configurations \textbf{[1,1,...,1]} are,
on average, very close in the three sets of workers, 7.04, 7.59 and
7.70 using vertical splitting and 7.94, 8.37 and 8.86 using horizontal
splitting. This clearly shows the limitations of the original stack
splitting approach when dealing with more workers not grouped in
teams. On the other hand, our platform scales better for the
configurations with more workers per team.


\section{Conclusions and Further Work}

We have presented a first implementation of an or-parallel Prolog
system specially designed to explore the combination of shared and
distributed memory architectures in clusters of multicores and we have
proposed a new set of built-in predicates that constitute the syntax
to interact with an or-parallel engine in our platform.

Experimental results showed that our platform is able to increase
speedups as we increase the number of workers per team, thus taking
advantage of the maximum number of cores in a machine, and to increase
speedups as we increase the number of teams, thus taking advantage of
adding more computer nodes to a cluster. We thus argue that our
platform is an efficient and viable alternative for exploiting
implicit or-parallelism in the currently available clusters of low
cost multicore architectures.

Despite these good results, our platform still suffers from
limitations that could be a basis to further research in the
area. Examples include: (i) support incremental copying between teams;
(ii) avoid speculative work; (iii) support dynamic code compilation;
and (iv) support dynamic teams. Further work also includes the ability
to pass specific information to the execution system about the
parallel computations at hand by using pre-defined directives. We may
have directives to define the execution model and the scheduling
strategy to be used, directives to define the number of workers,
directives to define granularity and load balancing policies, or
directives to define restrictions regarding speculative work and
predicate side-effects. We also plan to integrate our approach with
ThOr~\cite{CostaVS-10}, a thread-based extension of YapOr which uses
threads instead of processes to implement implicit or-parallelism, and
with another extension of YapOr which implements stack splitting on
shared memory~\cite{Vieira-12a}, thus allowing for the possibility of
using stack splitting to share work inside teams.


\section*{Acknowledgments}

This work was funded by the ERDF (European Regional Development Fund)
through the COMPETE 2020 Programme and by National Funds through FCT
(Portuguese Foundation for Science and Technology) as part of projects
UID/EEA/50014/2013 and ELVEN (POCI-01-0145-FEDER-016844) and through
the NORTE 2020 (North Portugal Regional Operational Programme) under
the PORTUGAL 2020 Partnership Agreement as part of project NanoSTIMA
(NORTE-01-0145-FEDER-000016). João Santos was funded by the FCT grant
SFRH/BD/76307/2011.


\bibliographystyle{acmtrans}
\bibliography{references}


\label{lastpage}
\end{document}